\documentclass[aps,nofootinbib,preprint,tightenlines,showpacs,showkeys]{revtex4}

\usepackage{epsfig,rotate}    
\usepackage{latexsym,color,citesort}

\usepackage{amssymb}
\usepackage{graphicx}

\usepackage{amsmath}

\usepackage{bm}

\newcommand{\beq}{\begin{eqnarray}}

\newcommand{\eeq}{\end{eqnarray}}
\newcommand{\eq}{\begin{eqnarray}}

\newcommand{\en}{\end{eqnarray}}

\newcommand{\bqa}{\begin{eqnarray}}

\newcommand{\eqa}{\end{eqnarray}}

\def\mqo2{{\!\!\!}}

\begin{document}

\preprint{HISKP-TH-10/16, FZJ-IKP(TH)--2010--12}

$\,$

\vspace{1.5cm}

\title{Extraction of the resonance parameters at finite 
times\footnote{Work supported in part by DFG (SFB/TR 16,
``Subnuclear Structure of Matter''), by the Helmholtz Association
through funds provided to the virtual institute ``Spin and strong
QCD'' (VH-VI-231) and by COSY FFE under contract 41821485 (COSY 106). 
We also acknowledge the support of the European
Community-Research Infrastructure Integrating Activity ``Study of
Strongly Interacting Matter'' (acronym HadronPhysics2, Grant
Agreement n. 227431) under the Seventh Framework Programme of EU.
A.R. acknowledges financial support
of the Georgia National Science Foundation (Grant \#GNSF/ST08/4-401).\\~}}

\author{Ulf-G. Mei{\ss}ner$^{1,2}$}\email{meissner@hiskp.uni-bonn.de}

\author{Kathryn Polejaeva$^1$}\email{polejaeva@hiskp.uni-bonn.de}

\author{Akaki Rusetsky$^1$}\email{rusetsky@hiskp.uni-bonn.de}

\affiliation{$^1$Helmholtz-Institut f\"ur Strahlen- und Kernphysik (Theorie),
and Bethe Center for Theoretical Physics, Universit\"at Bonn, 53115 Bonn, Germany}

\affiliation{$^2$Forschungszentrum J\"ulich, Institut f\"ur Kernphysik 
(Theorie), J\"ulich Center for Hadron Physics and Institute for Advanced 
Simulation (IAS-4), D-52425 J\"ulich, Germany
\vspace*{1.cm}}

\date{\today}

\begin{abstract}
In this paper we propose a model-independent method to extract 
the resonance parameters on the lattice directly from the Euclidean 
2-point correlation functions of the field operators at finite times. 
The method is tested in case of the two-point function of the 
$\Delta$-resonance, calculated at one loop in Small Scale Expansion.
Further, the
method is applied to a $1+1$-dimensional  model with two coupled 
Ising spins and  the results are compared with earlier ones obtained 
by using L\"uscher's approach.
\end{abstract}



\maketitle

\section{Introduction}
\label{sec:Intro}

It is well known that the asymptotic behavior of the two-point function
for large Euclidean times is determined by the lowest eigenvalue of the
Hamiltonian in a given channel. In case of stable particles, this property
allows one to determine their masses.
The case of the excited states is different. Here,
 the two-point function yields the spectrum of the
so-called ``scattering states.'' The relation to the energy and width of  
the resonance states is not direct, since a resonance, in general, can not
be associated with an isolated energy level of a Hamiltonian. 
Up to now, several alternative methods
have been used to determine these quantities from the lattice 
Monte Carlo (MC) simulations. These are:

\begin{itemize}
\item[i)]
At present, L\"uscher's approach~\cite{Houches,Wiese,Luscher:1990ux,signatures}
is widely used to deal with the resonances in lattice QCD, obtained in 
simulations with sufficiently low quark masses. In brief, the procedure
consists in determining first the phase shift by studying the volume dependence
of the energy spectrum on the lattice. Then, continuing the $S$-matrix into
the complex plane (e.g., by using the effective-range expansion whenever
possible), one attempts to determine the position of the poles on the second Riemann
sheet. This procedure is described in detail in Ref.~\cite{Hoja2}, where the
generalization to the case of the resonance matrix elements
(in 1+1 dimensions) is also considered. 
A shortcut is provided by using Breit-Wigner type 
parameterization for the scattering phase and  determining its parameters
(energy and width) from the lattice data (see, 
e.g. Ref.~\cite{Gockeler} and Refs.~\cite{Aoki:2007rd,Gockeler:2008kc}, 
where the method has been applied
in the case of the $\sigma$- and $\rho$-mesons, respectively). 
The L\"uscher's approach has been also generalized for the 
moving frames~\cite{Rummukainen}. 

\item[ii)]
Recently, the spectral functions in QCD have been reconstructed
 by using the maximal
entropy method (see, e.g.~\cite{Asakawa:2001,Sasaki:2002sj,Sasaki:2005ap}). 
This method, as well as L\"uscher's approach, has in principle the capacity to address
the problem of the extraction of the resonance energy and width from 
the Euclidean MC simulations on the lattice.

\item[iii)]
The Euclidean correlators have been parameterized in terms of 
the energy and width of an isolated resonance
state, in order to subsequently determine these quantities from the fit
to the lattice data~\cite{Michael}. In that paper, the method
has been applied to study the glueball decay. 

\item[iv)]
In certain cases, the decay width of an excited state can be evaluated by calculating decay
amplitudes on the lattice (see, e.g.~\cite{Loft:1988sy,Lellouch:2000pv}).

\item[v)]
Recently, there has been a substantial activity in the determination of the
excited meson and baryon spectrum by using generalized eigenvalue 
equations~\cite{Burch:2006cc,WalkerLoud:2008bp,Dudek:2009qf,Cohen:2009zk,Bulava:2009jb,Bulava:2010yg,Dudek:2010wm,Engel:2010my}. 
Despite spectacular progress achieved in the field, it should be stressed once again
 that a resonance state can not be uniquely associated with a particular 
energy level. To a certain extent, excited states and scattering states can
be distinguished, e.g., by studying the volume dependence of the spectral 
density~\cite{Alexandrou:2005gc,Alexandrou:2005ek}. This method, however
works for narrow resonances only~\cite{Niu:2009gt}.

\end{itemize}

In this paper, we combine some of the above ideas and propose
a systematic method to extract  resonance pole
positions from  lattice data. In its present form, our approach is applicable
to the systems with  a low-lying, well-isolated, narrow resonance in the 
spectrum (for example, the $\rho$ or the $\Delta\,(1232)$). 
First, we have tested our method using synthetic input data, represented
by the Euclidean propagator of the $\Delta$, 
calculated in the low energy effective field theory at one loop.
A further test has been carried out
in a 1+1 dimensional model of two coupled Ising spins, where the resonance
parameters have been determined in the past utilizing L\"uscher's 
approach~\cite{Gattringer:1991gp,Gattringer:1992yz}. In both cases, we find
that the method is capable to extract the pole position of the resonance.

The outline of the paper is as follows. In section~\ref{sec:kaellen} we discuss
the foundations of the method. The general representation of the two-point
function in the presence of a low-lying isolated resonance is discussed in
section~\ref{sec:representation}. In section~\ref{sec:fit} we consider the
procedure of the data fitting and the determination of the pole position by
using synthetic data. A short review of the 1+1 dimensional Ising model is given
in section~\ref{sec:ising}. The extraction of the resonance pole in this 
model is considered in section~\ref{sec:pole_ising}. Finally, 
section~\ref{sec:concl} contains our conclusions. Some technicalities are
relegated to the appendices.

\section{K\"allen-Lehmann representation}
\label{sec:kaellen}

\begin{figure}[t]
\begin{center}
\includegraphics*[width=6.cm]{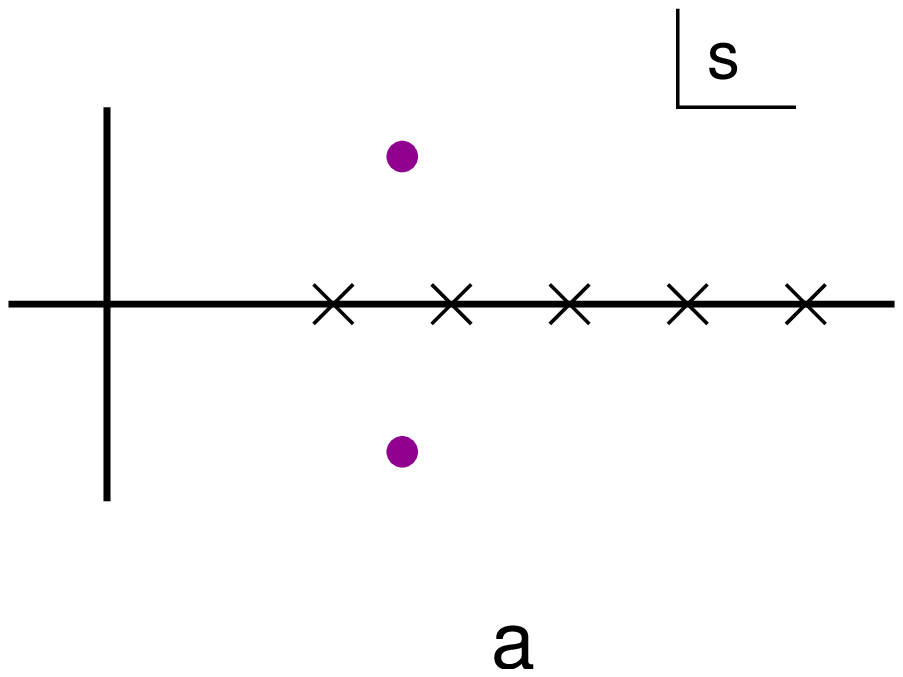}
\hspace*{1.cm}
\includegraphics*[width=6.cm]{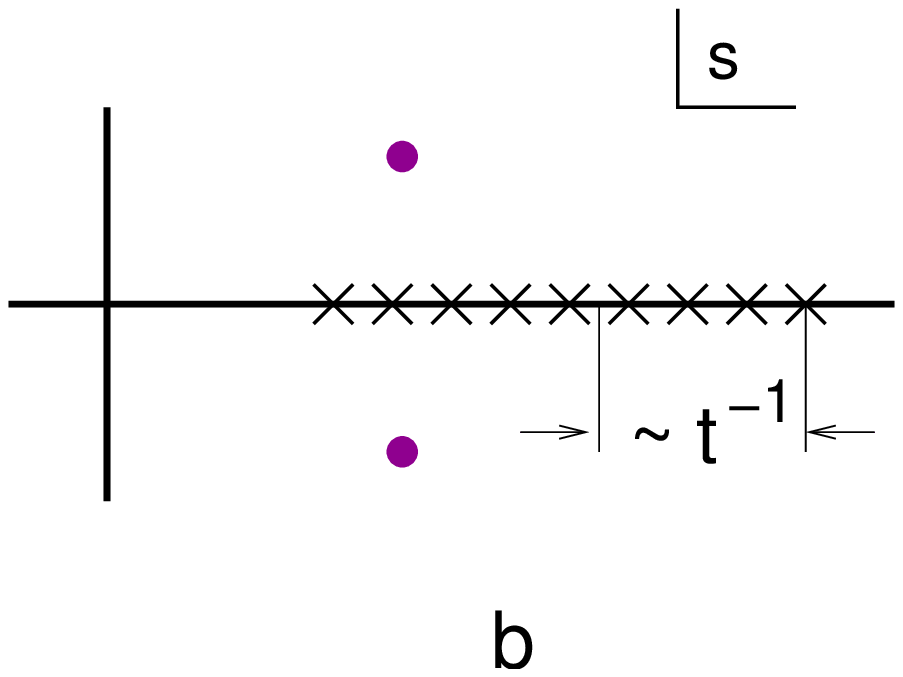}
\end{center}
\caption{Schematic representation of the analytic structure of the
  momentum-space two-point
  function in the rest frame $p_\mu=(\omega,0)$
as a function of the complex variable 
$s=(i\omega)^2$, for two different values of the box size $L$. The crosses 
on the real axis denote  the poles, and the shaded blobs, which
are located symmetrically to the real axis, mark the location of the resonance
poles on the second Riemann sheet emerging in the infinite volume limit.
At any finite value of $L$, however, the two-point function is meromorphic and
the second sheet does not arise. If the two-point function is
measured for Euclidean times $\leq t$, the ``energy resolution'' of such a
measurement is approximately equal to $t^{-1}$, as indicated on the right
 panel. }
\label{fig:plane}
\end{figure}

In the beginning of this section, we present a qualitative reasoning 
to justify our method. We start by mentioning that
 lattice QCD simulations are always carried out on  lattices 
with a finite (Euclidean) time and spatial extension. 
Below, we consider  lattices of the size
$T\times L^d$, where $T$ and $L$ denote the box size in time and space,
respectively, and $d$ is the number of spatial dimensions. For  not so 
large $L$ (however, large enough to suppress the 
polarization effects in stable particles), the energy levels are well separated
and  can in principle be extracted
 from the asymptotic behavior of the two-point function at large Euclidean
times $t\gg \Delta E$, where $\Delta E$ denotes the average level spacing.
The Fourier-transform of the
two-point function is a meromorphic function in the complex
$s$-plane 
(see Fig.~\ref{fig:plane}a). Consequently, the second Riemann sheet, as well
as the poles on it (corresponding to the resonances), do not appear at any
finite $L$. 
The information about these
 poles stays encoded in the dependence of the spectrum on the spatial box 
size $L$ and can be extracted (in several consecutive 
steps, as explained in the 
introduction) by using L\"uscher's approach.
 
In case an almost stable state is present, such a complicated procedure might
seem superfluous. Intuitively, it is clear that, if the decay width $\Gamma$
is small, the resonance will behave pretty much the same way as a
stable particle and will determine the $t$-dependence of the two-point
function in a large interval (however, not for the asymptotically large times
$t\gg\Gamma^{-1}$, when the resonance already has decayed). Consider, for example,
a lattice with the large spatial size $L$, and calculate the two-point function
at finite $t$ (see Fig.~\ref{fig:plane}b). If the ``resolution''
$\sim t^{-1}$ is larger than the distance between energy levels, the two-point
function is given by a sum of (many) exponentials with the spectral weight
suppressed by a factor $L^{-d}$, so that, effectively, the spectral sum 
transforms into an integral over the energies.
This is equivalent to the emergence of the cut
that connects the physical sheet to the second sheet. We expect that, in this case,
one may find an alternative representation of the two-point function in terms
of  {\em quasiparticle} degrees of freedom corresponding to the pole(s)
on the second Riemann sheet (i.e. the resonance 
decay and width) plus a small background which can be described by a 
few parameters. Namely, we expect that there exists a window in the 
variable $t$, where such a description will be more effective
than the multi-exponential representation through the
energies and spectral weights.

Hence, the original problem is reduced to finding  a universal,
model-independent parameterization of the two-point function in the presence
of a narrow resonance, which will allow one to determine the energy and the 
width
of the latter by performing a fit to the lattice data. This is analogous
to the exponential parameterization, which allows one to determine 
the mass of a stable particle by fitting at asymptotically large times. Note
that we shall try to avoid  approximations in the spectral function
(e.g. the narrow width approximation used in Ref.~\cite{Michael}), which are
a potential source of a systematic error. The contribution of the background,
albeit small, will be taken into account in a systematic manner.

The model-independent parameterization, which was mentioned above, 
can
be obtained directly 
by using the K\"allen-Lehmann representation for the two-point 
function. Below, we give a brief derivation of this representation in
a finite box. Further, we consider the limits $T\to\infty$ and $L\to \infty$
in detail, in order to quantify the qualitative arguments given in the 
beginning of this section.

In the derivation of the K\"allen-Lehmann representation in a finite volume,
we mainly follow the steps given in Ref.~\cite{Asakawa:2001}.
Let $\phi(x)\doteq\phi(t,{\bf x})$ be any local field,
carrying the resonance quantum numbers. The two-point function of this
field in the Euclidean space is defined as 
\eq\label{eq:twopoint}
D(t,{\bf x})&=&\langle N_t\,[ \phi(t,{\bf x})\phi^\dagger(0)]\,\rangle
=\frac{1}{Z(T)}\,\mbox{Tr}\,(N_t\,[\phi(t,{\bf x})\phi^\dagger(0)]\,
\mbox{e}^{-HT})~,
\nonumber\\[2mm]
N_t\,[\phi(t,{\bf x})\phi^\dagger(0)]
&\doteq&\theta(t) \phi(t,{\bf x})\phi^\dagger(0)
\pm\theta(-t)\phi^\dagger(0)\phi(t,{\bf x})\, ,\nonumber\\[2mm]
Z(T)&=& \mbox{Tr}\,(\mbox{e}^{-HT})\, ,
\en  
where the upper (lower) sign corresponds to bosons (fermions). 
The Fourier transform of this expression takes the form
\eq\label{eq:fourier}
D(t,{\bf x})=\frac{1}{TL^d}\sum_\omega\sum_{\bf k}D(i\omega,{\bf  k})\,
\mbox{e}^{-i\omega t-i{\bf k}{\bf x}}\, ,
\en
where $\omega=\frac{2\pi}{T}\, n_T$  $\bigl(\omega=\frac{\pi}{T}\, (2n_T+1)\bigr)$, with
$n_T\in \mathbb{Z}$,
are Matsubara frequencies in case of bosons (fermions), and
${\bf k}=\frac{2\pi}{L}\,{\bf n}$ with ${\bf n}\in \mathbb{Z}^d$.

Using a complete set  of  Hamiltonian eigenvectors 
$H|\alpha\rangle=E_\alpha |\alpha\rangle$ to calculate the trace in 
Eq.~(\ref{eq:twopoint}), one gets
\eq\label{eq:insert}
D(i\omega,{\bf  k})=-\frac{L^d}{Z(T)}\,\sum_\alpha\sum_\beta
\delta^d_{{\bf k},{\bf p}_\beta-{\bf p}_\alpha}\,
\frac{\mbox{e}^{-E_\alpha T}\mp\mbox{e}^{-E_\beta T}}
{E_\alpha-E_\beta+i\omega}\,
\langle\alpha|\phi(0)|\beta\rangle\langle\beta|\phi^\dagger(0)|\alpha\rangle
\, ,
\en
where $\delta^d_{{\bf k},{\bf q}}$ denotes the periodic
Kronecker $\delta$ in $d$ dimensions.
This relation can be rewritten as
\eq\label{eq:dispersion}
D(i\omega,{\bf  k})=\int_{-\infty}^\infty
\frac{d\omega'}{\omega'-i\omega}\,A(\omega',{\bf k})\, ,
\en
where the spectral function is given by
\eq\label{eq:spectral}
A(\omega',{\bf k})=\frac{L^d}{Z(T)}\,\sum_\alpha\sum_\beta
(\mbox{e}^{-E_\alpha T}\mp \mbox{e}^{-E_\beta T})\,
\delta(\omega'-E_\beta+E_\alpha)\,
\delta^d_{{\bf k},{\bf p}_\beta-{\bf p}_\alpha}\,
\langle\alpha|\phi(0)|\beta\rangle\langle\beta|\phi^\dagger(0)|\alpha\rangle
 .
\en
By applying discrete symmetries, it can be shown that the spectral function obeys
the following properties~\cite{Asakawa:2001}
\eq\label{eq:discrete}
A(\omega',{\bf k})\geq 0\quad\mbox{for}\quad \omega'\geq 0\, ,\quad\quad
A(-\omega',-{\bf k})=\mp A(\omega',{\bf k})=\mp A(\omega',-{\bf k})\, .
\en
The dispersion integral can be rewritten
as\footnote{For illustrative purpose, below we display the bosonic case only.
The fermionic case can be treated similarly.}
\eq\label{eq:dispersion-1}
D(i\omega,{\bf  k})= \int_0^\infty\frac{d{\omega'}^2}{{\omega'}^2+\omega^2}
\,A(\omega',{\bf k})\, .
\en
Finally, in the limit $T\to\infty$,  only the vacuum state $\alpha=0$,
$E_\alpha=0$ contributes, and
the spectral function in Eq.~(\ref{eq:dispersion-1}) is given by 
\eq\label{eq:analytic}
\lim_{T\to\infty} A(\omega',{\bf k})= L^d\,\sum_\beta
\delta(\omega'-E_\beta)\,
\delta^d_{{\bf k},{\bf p}_\beta}\,
|\langle 0|\phi(0)|\beta\rangle|^2\, .
\en
Now, 
let the variable $\omega^2$ be outside the narrow strip along the negative
real axis. Then, the function $2\omega'({\omega'}^2+\omega^2)^{-1}$ 
for all $0\leq \omega'<\infty$ is uniformly
bound from above by  some large constant $B$. For fixed $B$, performing the
 limit $L\to \infty$ and applying the regular summation
theorem~\cite{Luscher:1986pf},
 it is seen\footnote{The regular summation theorem
implies that the matrix elements $\langle 0|\phi(0)|\beta\rangle$ are 
continuous functions of $E_\beta$. We examine these matrix elements explicitly
in a simple quantum mechanical model in Appendix~\ref{app:separable} 
and show that in this case the above requirement is indeed fulfilled.} that the quantity 
$D(i\omega,{\bf k})$ approaches $D^\infty(i\omega,{\bf k})$, with the pertinent
spectral function given by
\eq\label{eq:infty}
A^\infty(\omega',{\bf k})=\int\!\!\!\!\!\!\!\!\!\sum_\beta
\delta(\omega'-E_\beta)\,
(2\pi)^d\delta^d({\bf k}-{\bf p}_\beta)\,
|\langle 0|\phi(0)|\beta\rangle|^2\, ,
\en
where $\int\!\!\!\!\!\!\sum_\beta$ 
stands for the sum (integral) over the continuous spectrum
wave functions. More precisely, for $\omega^2$ outside the strip
\eq\label{eq:cut}
|D(i\omega,{\bf k})-D^\infty(i\omega,{\bf k})|\leq B
\int_{\omega_{\sf min}}^\infty d\omega'|A(\omega',{\bf k})-A^\infty(\omega',{\bf k})|\, ,
\en 
where $\omega_{\sf min}$ is determined by the invariant mass of the lowest-mass
state. Further, according to the regular summation theorem, the difference
 $A(\omega_0,\Delta,{\bf k})-A^\infty(\omega_0,\Delta,{\bf k})$,
where
 \eq\label{eq:fromappendix}
A(\omega_0,\Delta,{\bf k})=\int_{\omega_0-\Delta/2}^{\omega_0+\Delta/2}
d\omega' A(\omega',{\bf k})\, ,\quad\quad
A^\infty(\omega_0,\Delta,{\bf k})=\int_{\omega_0-\Delta/2}^{\omega_0+\Delta/2}
d\omega' A^\infty(\omega',{\bf k})\, ,
\en
for any $\omega_0>\omega_{\sf min}$ and $\Delta>0$ 
converges faster than any 
power of $L$ as $L\to \infty$ (for a discussion, 
see Appendix~\ref{app:separable}). 
In other words, in this limit the 
two-point function converges to its infinite-volume counterpart everywhere
in the complex plane except the narrow strip along the cut
(for a related discussion, see also Ref.~\cite{DeWitt:1956be}). This 
statement is a mathematical formulation for the intuitive picture of
``poles merging into the cut,'' which is shown in Fig.~\ref{fig:plane}b. 
To further illustrate this, an example of a function which is meromorphic 
at a finite $L$ and develops a cut and a pole on the second Riemann sheet 
in the limit $L\to\infty$, is given
in Appendix~\ref{app:Herglotz}.

Moreover, from Eq.~(\ref{eq:fourier}) one finds
\eq\label{eq:t}
D(t,{\bf k})=\frac{1}{T}\,\sum_\omega \mbox{e}^{-i\omega t} D(i\omega,{\bf k})
&=&\int_{\omega_{\sf min}}^\infty d\omega'\,\frac{\mbox{e}^{-\omega'(T-t)}+\mbox{e}^{-\omega't}}
{1-\mbox{e}^{-\omega'T}}\,A(\omega',{\bf k})
\nonumber\\[2mm]
&\rightarrow& \int_{\omega_{\sf min}}^\infty d\omega' \mbox{e}^{-\omega't}A(\omega',{\bf k})\, .
\en
The last line is obtained in the limit $T\to\infty$. Together with the
expression~(\ref{eq:analytic}) for the spectral density, 
we recover the representation for $D(t,{\bf k})$ as a sum over exponentials.
In the case $t^{-1}$ is much larger than the distance between different 
energy levels (this can be achieved, e.g., by holding $t$ fixed and increasing 
$L$), many exponentials contribute to   $D(t,{\bf k})$ and the
sum over the energy eigenvalues can be replaced through the integral.
In this case, $A(\omega',{\bf k})$ is replaced by $A^\infty(\omega',{\bf k})$.

To summarize, the behavior of the two-point function can be 
studied in different regimes. 
For asymptotically large $t$ and moderately large $L$, 
only the few lowest, well-separated energy levels contribute. 
This situation is well described by a sum of a few exponential terms.
In difference to this, in the regime 
with asymptotically large $L$ and moderately large $t$
there are many terms
with nearly the same energies that contribute to the multi-exponential 
representation. The sum over the discrete energy spectrum effectively 
transforms into an integral. If, in addition, a low-lying well separated 
resonance emerges, we expect that the spectral integral can
be efficiently parameterized in terms of the resonance parameters instead
of the stable energy levels.

\section{Two-point function at finite  \boldmath $t$}
\label{sec:representation}

As discussed in the previous section, it is possible to perform the
infinite-volume limit $L\to\infty$ in the two-point function,
keeping the Euclidean time $t$ fixed. The spectral representation is given
by Eq.~(\ref{eq:dispersion-1}) with the spectral density given by 
Eq.~(\ref{eq:infty}). Note that the spectral density vanishes for
$0\leq \omega'\leq w_{\sf min}$, and hence the integration in 
Eq.~(\ref{eq:dispersion-1}) in fact is performed from 
${\omega'}^2=\omega_{\sf min}^2$ to
infinity.

For simplicity, we work in the center-of-mass (CM) frame ${\bf k}=0$
and denote $D(i\omega,{\bf 0})\doteq D(i\omega)$,
$A^\infty(\omega',{\bf 0})\doteq A(\omega')$. The spectral representation
then takes the form
\eq
D(i\omega)=\int_{\omega_{\min}}^\infty \frac{2\omega'd\omega'}
{{\omega'}^2+\omega^2}\,A(\omega')\, .
\en
In the vicinity of the elastic threshold, $A(\omega')
\sim (\omega'-\omega_{\sf min})^{l+1/2}$,
where $l$ stands for the orbital angular momentum\footnote{This statement 
is valid in 3+1 dimensions. In 1+1 dimensions, one has to 
substitute $l=0$ in all formulae.}.

Assume now that an isolated low-lying resonance emerges. This is equivalent
to the statement that the function $A(\omega')$ takes the form
\eq\label{eq:Az}
A(\omega')=\frac{(\omega'-\omega_{\sf min})^{l+1/2}}
{(\omega'-\omega_R)(\omega'-\omega_R^*)}\, Q(\omega')\, ,
\en
where the singularities of the function 
$Q(\omega')$ lie far enough 
from the threshold so that the Taylor
expansion of this function converges in the part of the complex plane
that includes the resonance poles at $\omega'=\omega_R$ and 
$\omega'=\omega_R^*$.
The energy and the width of the resonance is determined by $\omega_R$
in the  standard manner.
Note that these poles
come in  complex-conjugated pairs\footnote{$A(\omega')$ is the discontinuity of
a function which is analytic in the cut complex plane and obeys Schwarz reflection
principle. Hence the poles in this function (which emerge on the 
second Riemann sheet), always come in pairs. This is the justification
for the ansatz~(\ref{eq:Az}).}.

Using Eq.~(\ref{eq:Az}), it can be easily shown that
\eq
D(t)=\int_{\omega_{\sf min}}^\infty d\omega'\,\mbox{e}^{-\omega' t}
A({\omega'})=\mbox{e}^{-\omega_{\sf min}t}\,\int_0^\infty 
\frac{dE E^{l+1/2}\mbox{e}^{-Et}}{(E-E_R)(E-E_R^*)}\,
Q(E+\omega_{\sf min})\, ,
\en where
 $E_R=\omega_R-\omega_{\sf min}=E_0-i\Gamma/2$. 
As mentioned before, it is assumed that the Taylor expansion
$Q(E+\omega_{\sf min})
=\sum_{k=0}^{\infty} q_k E^k$ converges in the part of a
complex region which includes the resonance poles.

From the above expression we get
\eq
D(t)=\mbox{e}^{-\omega_{\sf min}t}\,\sum_{k=0}^\infty q_k 
F^{(k+l)}(t,E_R)\, ,\quad\quad
F^{(m)}(t,E_R)=\int_0^\infty 
\frac{dE E^{m+1/2}\mbox{e}^{-Et}}{(E-E_0)^2+\Gamma^2/4}\, .
\en
In particular, for $m=0,1$, we find
\eq
 \label{eq:F01}
F^{(0)}(t,E_R)&=&-\frac{2}{\Gamma}\, \mbox{Im}\,\chi(t,E_R)\nonumber\\[2mm]
F^{(1)}(t, E_R)&=&\mbox{Re}\,\chi(t, E_R)-\frac{2 E_{0}}{\Gamma}\,\mbox{Im}\,\chi(t,E_R),
\en
where
\eq
\label{eq:chi}
\chi(t,E_R)=
 \int_0^{\infty}  \frac{dE\,E^{1/2}\mbox{e}^{-Et}}{E-E_R}
\en
and the  following representation in form of an infinite series
is useful in a wide range of
the variable $t$
\eq\label{eq:chi1}
\chi(t,E_R)
=
-\pi\,\sqrt{-E_R}\, \mbox{e}^{-E_R t}
+ \sqrt{\frac{\pi}{t}}\,\biggl\{1+\sum _{k=0}^{\infty}\, 
\frac{(-2 E_R t)^{k+1}}{(2k+1)!!}\biggr\}\, .
\en
Further, the functions $F^{(m)}$ with $m\geq 2$ can be recursively
 expressed through $F^{(0,1)}$. The general representation of the two-point 
function follows straightforwardly from the above relations
\eq\label{eq:representation}
D(t)=\mbox{e}^{-\omega_{\sf min}t}\,\biggl\{
c_0 F^{(0)}(t,E_R)+c_1 F^{(1)}(t,E_R)
+\sum_{k=0}^\infty\frac{x_k}{t^{l+k+3/2}}\biggr\}\, ,
\en
where $c_{0,1}$ and $x_k$ are expressed through the Taylor coefficients $q_k$
as well as $E_0,\Gamma$.

Eq.~(\ref{eq:representation}) represents our central result.
It provides a universal parameterization of the Euclidean 
two-point function in the presence of a low-lying isolated resonance
described by two parameters $E_0,\Gamma$. The couplings $x_k$ are associated with
the non-resonant background. In particular, it encodes 
the contribution of the threshold
which lies {\em below} the resonance energy. This means that if $t$ is taken
too large, the background dominates and the information about $E_0,\Gamma$
is erased. We assume, however, that in the presence of a narrow resonance,
there exists a sufficiently wide window in $t$, where the background is small
and $E_0,\Gamma$ can be determined from the fit of the measured $D(t)$ to
the representation~(\ref{eq:representation}). We require that, in this window,
adding the background parameterized
by the constants $x_k$ should lead to small corrections in 
$E_0,\Gamma$ and the fit should remain stable 
against the increase of the number of independent $x_k$.

\begin{figure}[t]
\includegraphics[width=7.cm]{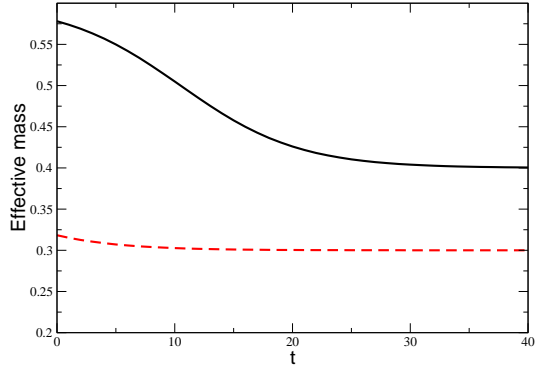}
\caption{Schematic representation of the effective mass for a stable particle
(dashed line) and for a resonance (solid line).
}
\label{fig:effmass}
\end{figure}

The physical meaning of our method can be easily illustrated by
Fig.~\ref{fig:effmass}. In this figure,
 the effective mass of a system in the presence
of a stable state/resonance is schematically depicted.
If there is  a stable particle,
the plateau in the effective mass sets in almost immediately.
However, if a resonance instead of a stable particle is present, there 
exists a wide window in $t$, where many excited states contribute and the
effective mass  decreases slowly until it reaches the asymptotic value. Our
method roughly 
corresponds to fitting $D(t)$ within this interval
 by the representation given in
 Eq.~(\ref{eq:representation}), and the decay width is
determined by the rate of the decrease of the effective
mass. Note that a similar picture was obtained in 
Ref.~\cite{Nolan:2006zz}. In that paper, the theory
of two coupled scalar fields, where the heavier field can decay into a 
couple of light scalars, was considered. In particular, it has been shown that
the effective mass of the heavy scalar, calculated on the lattice,
 exhibits the same behavior when
its mass is below/above the two-particle threshold 
(see Fig.~1 of that article). The two point functions of the excited mesons in
QCD also exhibit a similar behavior~\cite{Dudek:2009qf}.

It is instructive to compare the 
parameterization~(\ref{eq:representation}) to the pertinent formula obtained
in Ref.~\cite{Michael}. A technical difference consists in the absence of the
threshold factor $(\omega'-\omega_{\sf min})^{l+1/2}$ that results in a different
parameterization of the background. The important difference is, however, that
in Ref.~\cite{Michael} a Breit-Wigner type parameterization for the spectral
function is originally assumed. 
Since this ansatz did not fit the lattice data well, 
an {\em ad hoc} energy-dependence in the decay width has been introduced,
 and the functional form
of this dependence has been determined by using a trial-and-error method.
In our approach, the representation of $D(t)$, given in 
Eq.~(\ref{eq:representation}) is completely general and is based on
the sole assumption that a well-isolated resonance emerges at low 
energies. The background is parameterized by the constants $x_k$ in a 
systematic manner.

\section{The fit}
\label{sec:fit}

We first test our method by using synthetic data. Consider, for instance, the
propagator of the $\Delta$-resonance evaluated 
in the Small Scale Expansion (SSE)\footnote{The SSE is a phenomenological extension
of Chiral Perturbation Theory in which the $\Delta$-nucleon mass splitting
is counted as an additional small parameter. This quantity, however, does not
vanish in the chiral limit. The framework of the SSE is laid in detail in 
Ref.~\cite{Hemmert:1997ye}.} 
at one loop~\cite{Bernard:2007cm,Bernard:2009mw}.
In the Minkowski space, this propagator is given by
\beq\label{eq:SD-1}
S_\Delta(p)&=&-S_\Delta^{3/2}(p)\Pi^{3/2}+\mbox{spin-1/2}\, ,\nonumber\\[2mm]
S_\Delta^{3/2}(p)&=&\frac{1}{\mathring{m}_\Delta(1+\Sigma_2(p^2))-\not\! p(1-\Sigma_1(p^2))}\, ,
\eeq
where $\mathring{m}_\Delta$ denotes the mass of the $\Delta$-resonance 
in the chiral limit, $\Pi^{3/2}$ stands for the projector onto the 
spin-3/2 state and the spin-1/2 part does not have a pole in the 
low-energy region (for a general proof of this statement, see \cite{Krebs:2009bf}). 
Further, the invariant functions $\Sigma_{1,2}(p^2)$
at order $p^3$ in the chiral expansion are given by
\eq\label{eq:SD-2}
\Sigma_1(s)=\frac{c_A^2}{F^2}\,(W_2(s)-W_3(s))\, ,\quad\quad
\Sigma_2(s)=\frac{c_A^2}{F^2}\,\frac{m_N}{m_\Delta}\, W_2(s)\, ,
\en
where $c_A$ and $F$ denote the $\pi N\Delta$ coupling constant and the pion
decay constant in the chiral limit, respectively, $m_{N},m_\Delta,M_\pi$ are the 
nucleon, $\Delta$ and pion masses, respectively,
 and the invariant functions $W_{2,3}$ are given by~\cite{Bernard:2007cm,Bernard:2009mw}
\eq\label{eq:SD-3}
W_3(s)&=&\frac{-s+m_N^2-M_\pi^2}{-2s}\,W_2(s)
+\frac{M_\pi^4}{128\pi^2s}\,\biggl( \ln\frac{M_\pi^2}{m_N^2}
+\frac{1}{6}\biggr)\, ,
\nonumber\\[2mm]
W_2(s)&=&-\frac{1}{12s} \,\biggl(\lambda W_0(s)-(-s+m_N^2-M_\pi^2)\frac{M_\pi^2}{16\pi^2}\,\ln\frac{M_\pi^2}{m_N^2}\biggr)\, ,
\nonumber\\[2mm]
W_0(s)&=&\frac{i\sqrt{\lambda}}{16\pi s}
-\frac{s-m_N^2+M_\pi^2}{32\pi^2s}\,\biggl(\ln\frac{M_\pi^2}{m_N^2}-1\biggr)
-\frac{\sqrt{\lambda}}{32\pi^2s}\,
\ln\frac{s+M_\pi^2-m_N^2+\sqrt{\lambda}}{s+M_\pi^2-m_N^2-\sqrt{\lambda}}\, ,
\nonumber\\[2mm]
\lambda&=&(s-(m_N+M_\pi)^2)(s-(m_N-M_\pi)^2)\, .
\en
The trace of $S_\Delta^{3/2}(p)$ obeys the dispersion relation
\eq\label{eq:SD-4}
\mbox{Tr}\,S_\Delta^{3/2}(p)=\frac{4\mathring{m}_\Delta(1+\Sigma_2(p^2))}
{(\mathring{m}_\Delta(1+\Sigma_2(p^2)))^2-p^2(1-\Sigma_1(p^2))^2}
=\int_{(m_N+M_\pi)^2}^\infty\frac{ds'}{s'-p^2-i\epsilon}\,A(s')\, ,
\en
where the expression for the discontinuity can be directly read off
from Eqs.~(\ref{eq:SD-1})-(\ref{eq:SD-3}).

In the calculations we have used the following values of the parameters:
$m_N=940~\mbox{MeV}$, $M_\pi=140~\mbox{MeV}$, 
$m_\Delta=\mathring{m}_\Delta=1232~\mbox{MeV}$, $F=F_\pi=92.4~\mbox{MeV}$ and $c_A=1.5$ 
(this value leads to the width $\Gamma=124~\mbox{MeV}$ in a $O(\epsilon^3)$
calculation at $p^2=m_\Delta^2$). It is easy to check that the propagator has a pole at 
$m_R=1212~\mbox{MeV}$ and $\Gamma=76~\mbox{MeV}$ (note the large shift
in the quantity $\Gamma$ as compared to its value obtained at 
$p^2=m_\Delta^2$ that presumably is an artefact of a
$O(\epsilon^3)$ approximation).

Next, we wish to investigate whether it is possible to recover this result
by applying our method. To this end, we analytically continue 
Eq.~(\ref{eq:SD-4}) into  Euclidean space and perform the Fourier transform
with respect to the fourth component of the momentum. 
The resulting values are treated as synthetic data. We choose the interval
$1.7 M_\pi^{-1}<t<4M_\pi^{-1}$ and perform a  least squares fit of these
data to the formula~(\ref{eq:representation})
 (the data points are assumed to be distributed 
equidistantly in this interval). 

In the fit, we cut the sum in Eq.~(\ref{eq:representation}) at some
value $k_{\sf max}$. The fit of the 7 data points with $k_{\sf max}=0$
yields  $m_R=1213~\mbox{MeV}$ and $\Gamma=74~\mbox{MeV}$ that is
already close to the exact values. The procedure converges rapidly.
At the accuracy of the digits displayed, 
the exact result is obtained for $k_{\sf max}=2$.
Adding more terms, it is possible to improve the agreement with the exact
result up to very many decimal digits.

To summarize, using synthetic data, we have demonstrated that our method is
capable to reconstruct the exact position of a pole in a complex plane 
from a limited data sample. To perform a similar analysis
for  real Monte Carlo data is much more challenging.
 One of the main problems that we have encountered there,
is related to the instability of the fit when $k_{\sf max}$ increases 
(this problem
already arises for relatively small $k_{\sf max}=3~\mbox{or}~4$).
Namely, the constants $x_k$, which describe the background,
become very large in magnitude having alternating signs and 
this destabilizes the values of $E_0,\Gamma$ extracted from the fit.

In order to circumvent this problem, we have performed a Bayesian fit to the
lattice MC data. A detailed description of the Bayesian fit techniques,
which is well suited for our purposes, can be found, e.g. in 
Ref.~\cite{Schindler:2008fh}. We shall present a brief summary of the 
method below. The function to be minimized in the standard least 
squares fit is given by
\eq\label{eq:chi2}
\chi^2=\sum_i(D(t_i,E_0,\Gamma,c_1,c_2,x_k)-\bar D(t_i))^2\, ,\en
where $\bar D(t_i)$ are data corresponding to the points $t_i$.
In Eq.~(\ref{eq:chi2}) it is implicitly assumed that the MC errors in the data
$\bar D(t_i)$ do not vary much with $t_i$. Note that
the above form still does not include our prior knowledge about $x_k$.
The assumption about the smoothness of the function $Q(\omega')$ in 
Eq.~(\ref{eq:Az}) implies that $x_k$ should be
 of ``natural size''  excluding the
scenario where the $x_k$ become large with alternating signs.

In order to implement this prior knowledge into the fitting procedure, 
in analogy with Ref.~\cite{Schindler:2008fh},
we define the augmented $\chi^2$
\eq\label{eq:aug}
\chi^2_{\sf aug}=\chi^2+\chi^2_{\sf prior}\, ,\quad\quad
\chi^2_{\sf prior}=\frac{1}{S^2}\,\sum_{k=0}^{k_{\sf max}}x_k^2\, ,
\en
where $S$ is some scale that ensures that all $x_k$ stay in the ``natural'' 
range.

We determine the quantity $S$ by using the trial-and-error method. If $S$ is too
large, the introduction of $\chi^2_{\sf aug}$ does not cure the problem with
the convergence. This sets the upper limit on the value of $S$. The lower limit
for $S$ is set by the requirement that the results obtained with standard
$\chi^2$ and $\chi^2_{\sf aug}$ agree for low $k_{\sf max}=1,2$. In addition,
within
this range, the final result of the fit for $E_0,\Gamma$ should not depend
on $S$.

In the 1+1 dimensional model with two Ising spins discussed in the next
section, we have performed  
fits using $\chi^2_{\sf aug}$. Below we show that this technique allows one 
to extract the precise values of $E_0,\Gamma$ from the lattice MC data
in this model.

\section{$1+1$ dimensional model with two coupled Ising spins}
\label{sec:ising}

In this section we apply our method to
 the extraction of the resonance pole position
to a 1+1 dimensional model of two coupled Ising spins. This model has been
treated in  Refs.~\cite{Gattringer:1991gp,Gattringer:1992yz}
using L\"uscher's approach. In particular, it has been shown that a narrow 
resonance emerges in the system, whose  parameters can be
 extracted in a systematic manner.

The action of the model is given by
\eq\label{e36}
S= -\kappa_{\phi} \sum_{z\in\Lambda,\hat{\mu}=1,2} \phi_{z}\, 
\phi_{z+\hat{\mu}} -\kappa_{\eta} \sum_{z\in\Lambda,\hat{\mu}=1,2}
 \eta_{z}\eta_{z+\hat{\mu}}+ \frac{g}{2}\sum_{z\in \Lambda,\hat{\mu}=1,2}
\eta_{z}\phi_{z} \, (\phi_{z-\hat{\mu}}+\phi_{z+\hat{\mu}}),
\en
where $\phi_z,\eta_z=\pm 1$ are two Ising spins which interact with each
other through the Yukawa-type coupling $g\eta\phi\phi$. 
The sum $z\in \Lambda$, where
$z=(x,t)$, runs 
over all lattice points and $\hat \mu$ denotes the unit vector along 
the spatial axis. The couplings 
$\kappa_{\phi},\kappa_{\eta}>0$ are chosen so that the masses of $\phi$ and
$\eta$  are $m_{\phi}\simeq 0.19$ and $m_{\eta}\simeq 0.5$ (in lattice units).
Note that, if $g\neq 0$, the $\eta$ decays into $2\phi$, so $m_\eta$ corresponds
to the resonance energy in this case.

The model has been analyzed in detail in 
Refs.~\cite{Gattringer:1991gp,Gattringer:1992yz}. We give only a short
summary of this analysis here. In particular, it has been argued that in the
theory described by the Lagrangian~(\ref{e36}) no second-order phase transition
occurs and thus the continuum limit can not be performed. In other words, 
all results obtained here refer to the effective theory with an ultraviolet
cutoff.

The energy spectrum is determined by solving the
 generalized eigenvalue problem. The operator basis is defined as in 
Refs.~\cite{Gattringer:1991gp,Gattringer:1992yz}
\eq
\mathcal{O}_1(t)=\frac{1}{L}\,\sum_x \eta_{x,t}\, ,\quad\quad
\mathcal{O}_j(t)=\frac{1}{L^2}\,\sum_{xy} \phi_{x,t}\phi_{y,t}
\mbox{e}^{-ip_j(x-y)}\, ,\quad p_j=\frac{2\pi(j-2)}{L}\, ,
\en
with $j=2,3,\cdots$. The correlator matrix is given by
\eq
 \label{e38}
C_{ij}(t)=\langle\mathcal{O}_{i}(t)\mathcal{O}_{j}(0)\rangle  -\langle\mathcal{O}_{i}(t)\rangle \langle \mathcal{O}_{j}(0)\rangle\, .
\en
The spectral decomposition of $C_{ij}(t)$ is approximated by the 
truncated series
\eq
C_{ij}(t)=\sum_{l=1}^rv_i^{(l)}{v_j^{(l)}}^*\,\mbox{e}^{-W_lt}\, .
\en
The energy eigenvalues $W_l$ for $l=1,\cdots , r$ are determined by diagonalizing
the matrix $M(t,t_0)\doteq C^{-1/2}(t_0)C(t)C^{-1/2}(t_0)$, where $t_0$ is some fixed
time (in the following, as in Refs.~\cite{Gattringer:1991gp,Gattringer:1992yz},
we always use $t_0=1$). The eigenvalue equation takes the form
\eq
M(t,t_0)u^{(l)}=\lambda^{(l)}(t,t_0)u^{(l)}  \, ,\quad\quad
\lambda^{(l)}(t,t_0)=\mbox{e}^{-W_l(t-t_0)}\, ,\quad l=1\cdots r\, ,
\en
where $u^{(l)}$ form an orthonormal  basis.
The eigenvectors $v^{(l)}$ are given by 
$v^{(l)}=\exp(W_lt/2)\,C^{1/2}(t_0)u^{(l)}$.

The MC simulation is done by using a cluster 
algorithm~\cite{SwendsenWang:1987}. We closely followed the procedure 
described in~\cite{Gattringer:1991gp,Gattringer:1992yz} and, using the
parameter set $\kappa_\phi=0.3700$, $\kappa_\eta=0.3700$, $g=0.04$,
have reproduced the $L$-dependent spectrum calculated in this paper. The
resonance parameters found in 
Refs.~\cite{Gattringer:1991gp,Gattringer:1992yz} are: $m_\eta=0.5112(3)$
and $\Gamma_\eta=0.0100(3)$.
It remains to be seen, whether the same result can be obtained by 
using our approach.

\section{Results}
\label{sec:pole_ising}

\begin{figure}[t]
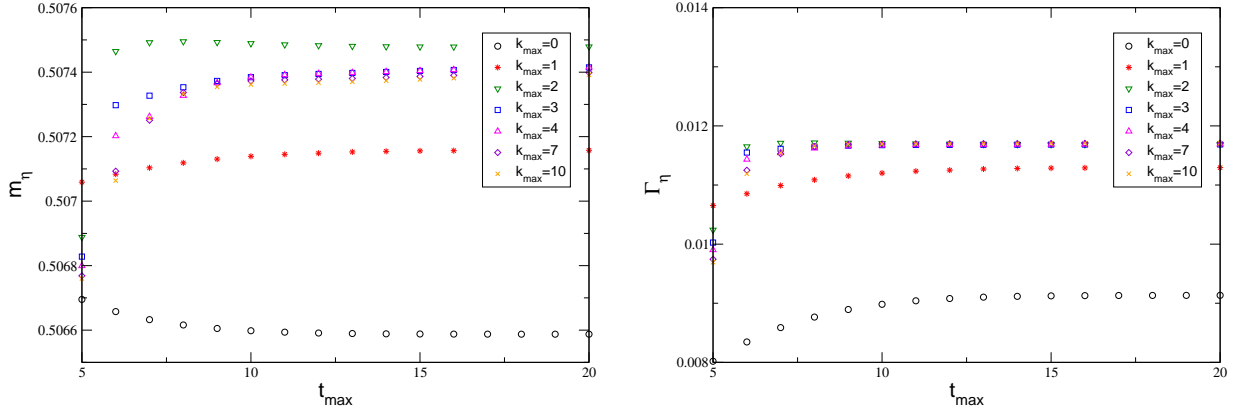

\includegraphics[height=5.4cm]{E-conv_60.eps}\hspace*{.48cm}
\includegraphics[height=5.4cm]{G-conv_60.eps}
\caption{Checking the convergence of $m_\eta$ (left panel) and $\Gamma_\eta$ (right panel) 
against the variation of $t_{\sf max}$ and $k_{\sf max}$, for $L=60$.
A similar behavior is observed for smaller values of $L$.}
\label{fig:tmax-kmax}
\end{figure}

In order to use our method, one has to calculate the two-point function
within a sufficiently large interval in the Euclidean time $t$ and then
fit the result with Eq.~(\ref{eq:representation}). To this end,
the correlator $C_{11}(t)$ has been chosen, see Eq.~(\ref{e38}). However,  
as already mentioned in~\cite{Gattringer:1991gp,Gattringer:1992yz}, 
the simulations become unstable already at $t\simeq t_{\sf unst}= 5-8$, 
depending on the value of $L$ chosen. The statistical error in the   
effective mass of $\eta$ at $t>t_{\sf unst}$
blows up, rendering an accurate fit impossible.
The use of improved estimators or a substantial increase of the 
number of configurations results only in a moderate improvement
of the error in $C_{11}(t)$.

As described above, 
the energy spectrum of the system can be determined with high 
accuracy from the correlator matrix $C_{ij}(t)$ at $t\leq t_{\sf unst}$ by 
applying the generalized eigenvalue method. 
In addition to the ground state, the approach allows a reliable extraction
of higher excited levels (up to 4 or 5 levels, depending on $L$).
The physical reason for this is
that the matrix $C_{ij}(t)$ contains much more information about the system
than the single function $C_{11}(t)$. In particular, it contains
information about the matrix elements describing the transitions between 
various energy levels.
 
So, it is not surprising  that using this input in our method
helps to  reduce the errors dramatically and to stabilize the fit. In brief,
the procedure can be described as follows:
\begin{enumerate}
\item
The energy spectrum $W_l$ and the wave functions $v^{(l)}$ are accurately
determined by measuring the matrix $C_{ij}(t)$ at $t\leq t_{\sf unst}$.
We choose $t_{\sf unst}=5$ for all $L$ and average all $W_l$
for $t=t_0\cdots (t_{\sf unst}-1)$.  
\item
The function $C_{11}(t)$ is approximated by the multi-exponential function
$C_{11}(t)=\sum_{l=1}^rz_l\exp(-W_lt)$, where the $z_l=|v_1^{(l)}|^2$ are averaged
for all $t=t_0\cdots (t_{\sf unst}-1)$. This approximation
is used for $t> t_{\sf unst}$ as well.
Note that $z_l,~l\neq 1$ encode the information about the overlap of $\eta$ and
$2\phi$ states that determines the decay width of a resonance. 
\item
The expression~(\ref{eq:representation}) is fitted to the $C_{11}(t)$ which is approximated
by the multi-exponential function.
\end{enumerate}
The MC simulations were carried out for various lattice sizes in the interval 
$L=24-60$, while the value $T=100$ remained
 fixed throughout the simulations. We have
used bases containing 4-6 operators and performed test runs for some (large)
values of $L$ by using the basis of 8 and 10 
operators. In the fit, all data between $t=1$ and $t=t_{\sf max}$ were
used. The errors in our results are purely 
statistical and were estimated by performing 5 independent simulations
with $10^{6}$ configurations each. In addition, we find that
the increase of the number of operators to 8 or 10
operators does not affect the result within the errors.

First, the stability of our results was checked, when
 $t_{\sf max}$ and $k_{\sf max}$ increase. The result of this check  
 is displayed in Fig.~\ref{fig:tmax-kmax}, where the dependence 
of the real and imaginary parts of the resonance pole position on $t_{\sf max}$
is plotted for different values of $k_{\sf max}$.
It is seen that for $t_{\sf max}\geq 10-12$ 
both the energy and the width remain almost constant
 and converge rapidly in $k_{\sf max}$ already at 
$k_{\sf max}=3$, if the Bayesian fit is performed. The similar behavior 
is observed at all values of $L$.
The final result for the resonance pole parameters 
is always given at $k_{\sf max}=10$.

\begin{figure}[t]

\includegraphics[width=11.cm]{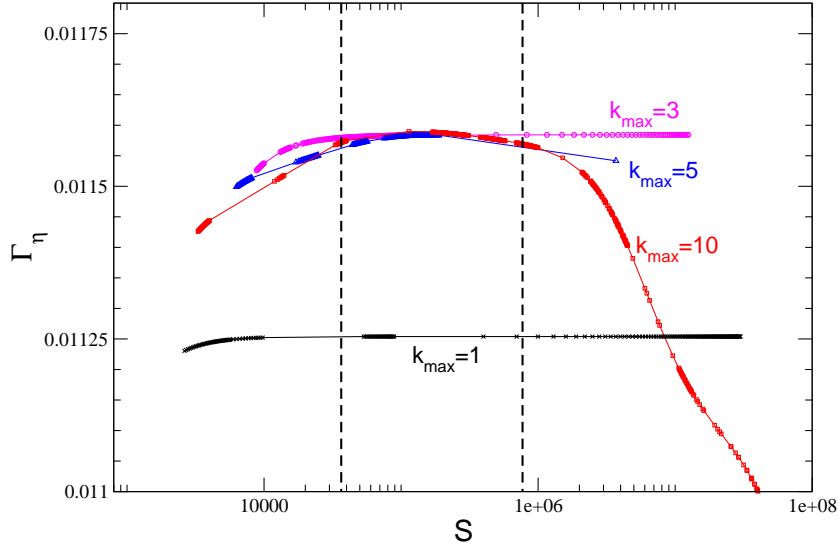}
\caption{Dependence of the width on the scale $S$ used in the Bayesian fit at
different $k_{\sf max}$. For small values of $S$, the result is
scale-dependent. For large $S$, the result does not converge with 
$k_{\sf max}$. There exists a plateau around $S\simeq 10^5$ where the
procedure converges and yields a scale-independent result.}
\label{fig:S}
\end{figure}

In order to ensure that, performing the Bayesian fit, 
a bias is not introduced in the extracted values of the 
resonance parameters, one has to check
that there exists a range of the scale parameter $S$
where the energy and the width depend weakly on $S$.
The results for both quantities at different values of $L$ look qualitatively
similar. In Fig.~\ref{fig:S} we present the plot for the width at $L=60$. 
As seen from Fig.~\ref{fig:S},  a wide plateau emerges around
$S\simeq 10^5$, where the scale dependence practically disappears while the 
convergence in $k_{\sf max}$ still persists. This is the window, where
the extraction of the width is finally
carried out. Increasing $S$ even further, the
convergence in $k_{\sf max}$ breaks down, and the result can not be trusted
any longer.

\begin{figure}[t]
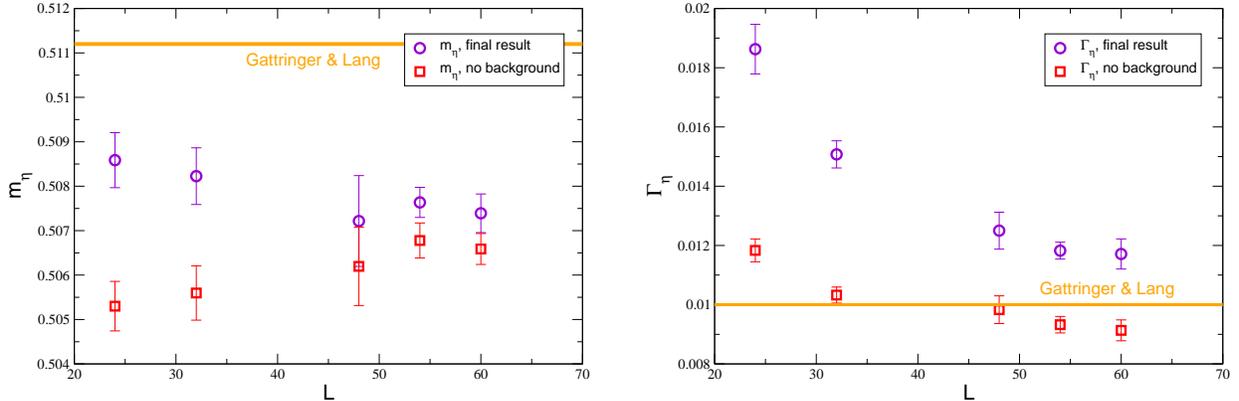

\includegraphics[height=5.3cm]{E-L.eps}\hspace*{.6cm}
\includegraphics[height=5.3cm]{Gamma-L.eps}
\caption{The energy and the width of the resonance, extracted from the data
at $k_{\sf max}=10$ (final result).
For comparison, we display the result at $k_{\sf max}=0$ (no background) and
the result, taken from Refs.~\cite{Gattringer:1991gp,Gattringer:1992yz}
(the error quoted in these references corresponds to the thickness of the lines).
The errors in our calculations are purely statistical.}

\vspace*{1.1cm}

\label{fig:L}
\end{figure}

Finally, since our MC data have been calculated at a finite $L$, whereas
 the formula~(\ref{eq:representation})
 refers to the limit $L\to\infty$, there is  an expected residual 
volume dependence in the parameters $E_0,\Gamma$.
The Fig.~\ref{fig:L} displays this dependence.
In particular, it is seen that there is a rather strong variation
of the width at small values of $L$ that flattens around $L=48$.
In the present paper we do not attempt to
quantitatively describe the finite volume artefacts.
This issue forms the subject of a separate investigation and
we plan to address it in the future. 

From Fig.~\ref{fig:L} it is also seen that the effect of the background 
on the real part of the pole position is small, whereas the imaginary 
part is far more sensitive to it. Namely, the two values of $E_0$, calculated
at  $L=60$ for $k_{\sf max}=0$ and  $k_{\sf max}=10$ differ by $\simeq 0.2\%$,
whereas the same calculation for the width yields
 $\Gamma_\eta=(0.91\pm 0.04)\cdot 10^{-2}$ and
 $\Gamma_\eta=(1.17\pm 0.05)\cdot 10^{-2}$, respectively.
In general, one may conclude that the effect of the background can not
be neglected.

\begin{figure}[t]

\includegraphics[width=11.cm]{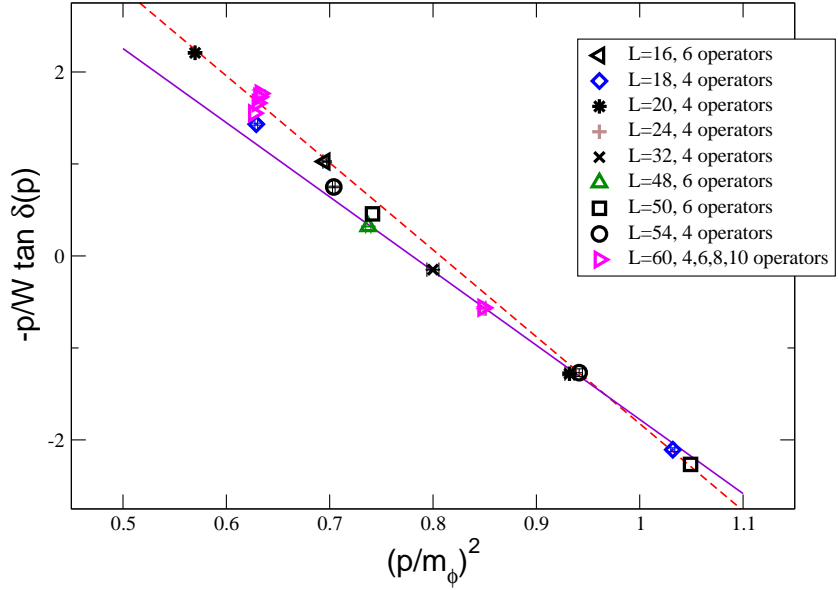}
\caption{The quantity $-p/W\,\tan\delta(p)$, extracted from the energy 
spectrum, vs. $(p/m_\phi)^2$. The solid line is the linear fit $a-bp^2$, with
$a,b$ determined from Eqs.~(\ref{eq:final}) and (\ref{eq:Rp}), using the
central values of $m_\eta$ and $\Gamma_\eta$. To draw the dashed line, we have 
used $m_\eta,\Gamma_\eta$ from Ref.~\cite{Gattringer:1991gp} instead
of Eq.~(\ref{eq:final}) (cf. with Fig.~2 of that paper). The errors in the
data are purely statistical. In addition, we have indicated the dimension of 
the operator basis $O_j$ used to extract the spectrum.}
\label{fig:luescher}
\end{figure}

The final result for the real and imaginary parts of the pole position
(for $L=60$) are
\beq\label{eq:final}
m_\eta &=& 2m_\phi+E_0=0.5074\pm 0.0004\, ,\nonumber\\[2mm]
\Gamma_\eta &=& \Gamma=(1.17\pm 0.05)\cdot 10^{-2}
\eeq
 (errors are only statistical). 
This result can be checked by using
the effective-range expansion for the scattering phase 
(cf. with Ref.~\cite{Gattringer:1991gp})
\eq\label{eq:effrange}
-\frac{p}{W}\,\tan\delta(p)=a-bp^2\, ,\quad\quad
W=2\sqrt{m_\phi^2+p^2}\, ,
\en
where the parameters $a$ and $b$ are related to $m_\eta$ and $\Gamma_\eta$ through
\eq\label{eq:Rp}
m_\eta=2\sqrt{m_\phi^2+\frac{a}{b}}\, ,\quad\quad
\Gamma_\eta=\frac{4}{bm_\eta^2}\sqrt{\frac{a}{b}}\, .
\en
As one sees from Fig.~\ref{fig:luescher}, our phase shift results
are generally in 
agreement with the results of the Ref.~\cite{Gattringer:1991gp}.
However, since the data are not {\em exactly} linear, 
the question arises, which interval in the variable $p^2$ should be used in the
fit to determine the coefficients $a$ and $b$. For instance, 
the extracted values of the phase shift
{\em in the vicinity of the resonance} $(p/m_\phi)^2\simeq 0.8$ 
neatly follow the straight line with the parameters $a,b$, which were
determined from Eqs.~(\ref{eq:effrange}), using the central values of
$m_\eta,\Gamma_\eta$ in Eq.~(\ref{eq:final}).

Now, we are in a position to compare our results to those of 
Refs.~\cite{Gattringer:1991gp,Gattringer:1992yz}.
The difference in the
real part of the resonance pole position is small -- both results agree with an 
accuracy of better than one per cent. 
The effect is larger in the imaginary
part. However, one should keep in mind that the magnitude of the imaginary 
part is approximately 50 times smaller than the real part.
As one concludes from Fig.~\ref{fig:luescher}, a relatively
 large effect in the imaginary part could be
related, e.g., to the fact that the effective 
range plot is not exactly linear. Therefore, it
seems plausible that the systematic errors both in 
Refs.~\cite{Gattringer:1991gp,Gattringer:1992yz} and in the present paper
are underestimated. We expect that the results should agree within the errors.

Comparing our method to  L\"uscher's approach, we further note that, 
once the plateau in $L$ sets in,
the energy and the width within our method
can be extracted at {\em a single} value of $L$. 
In contrast to this, 
L\"uscher's approach implies the study of the volume dependence of the
energy levels. This difference can be related to the fact that our method
uses additional input information from MC simulations.
In particular, apart from the energy spectrum, the
 two-point function $C_{11}(t)$ contains the information about the transition
 matrix elements encoded in the constants $z_l,~l\neq 1$.

Last but not least, we have also checked that our method works in the 
non-interacting case as well. Setting $g=0$ and adjusting 
$\kappa_\phi,\kappa_\eta$ in the Lagrangian to keep the masses of 
$\phi$ and $\eta$ the same as in the interacting case (see 
Refs.~\cite{Gattringer:1991gp,Gattringer:1992yz}), 
we have done the calculation of the function $C_{11}(t)$ anew.
The fitted width turns out to be two orders
of magnitude smaller as compared to the interacting case. This obviously
corresponds to a stable particle.

\section{Conclusions}
\label{sec:concl}

\begin{itemize}

\item[i)]
In the present paper we have proposed a novel method to extract 
the resonance pole position on the 
lattice. The method is based on the universal representation of the
two-point function Eq.~(\ref{eq:representation}), which is valid given the sole
assumption that an isolated low-lying resonance is present in the system.
The energy and the width of this resonance are determined from the fit of 
Eq.~(\ref{eq:representation}) to the lattice MC data. It remains to be seen
whether such a universal representation can be derived in more complicated
cases (e.g., for the multi-channel scattering, see Ref.~\cite{Lage:2009zv}) as well.

\item[ii)]
The proposed
 method provides an alternative to  L\"uscher's approach to the resonances.
In the latter, the {\em volume dependence} of the {\em spectrum} on the
moderately large lattices is studied. 
The spectrum consists of the scattering states only -- 
the resonance has already decayed.  In our approach the
two-point function is studied 
{\em at  finite times $t$} (when the resonance is still
``alive'') and for the {\em asymptotically large} values of $L$. Note that
the actual calculations do not seem to require
extraordinarily large volumes. For example, in the 1+1 dimensional Ising model
 $L=48$ was already
sufficient.

\item[iii)]
The above difference entails an important advantage of the method described
in this paper:
whereas in L\"uscher's approach the MC simulations should be performed at least
at several volumes in order to extract the resonance, the measurement
at one, albeit sufficiently large, lattice volume suffices in our method.

\item[iv)]
In certain cases,
the numerical accuracy of the method can be improved considerably, 
if the multi-exponential
representation of the two-point function is used in the fit
instead of calculating this function
directly through MC simulations
at all values of $t$.
 The coefficients of the multi-exponential
 representation are obtained by solving the
generalized eigenvalue problem and,  in particular, encode
the decay matrix elements.

\item[v)]
Recently, the excited meson and baryon spectra have been determined by 
several lattice collaborations 
using the generalized eigenvalue equation (see, e.g. 
Refs.~\cite{Bulava:2010yg,Dudek:2010wm,Engel:2010my} for the latest work on the subject).
These calculations closely resemble the calculations in the 1+1 dimensional
toy model, which were presented in this paper. In our opinion, it would be
very interesting to apply the proposed method to the data 
and if possible try to locate
the resonance pole(s). This can be done at {\em no additional cost,}
 since the
results of already existing MC simulations would be used.

\end{itemize}

\begin{acknowledgments}
The authors are thankful to J\"urg Gasser for a close and fruitful cooperation
at all stages of the work on the project. We wish to thank
Ferenc Niedermayer for his constant
readiness to help, for numerous discussions and suggestions.
We also thank R. G. Edwards, C. Gattringer,
M. G\"ockeler, P. Hasenfratz, C. Lang, 
D. Lee, C. McNeile, C. Michael, D. Phillips, 
M. Schindler, K. Urbach and U. Wenger for interesting discussions.

\end{acknowledgments}

\appendix

\section{Continuum limit in the matrix elements}
\label{app:separable} 

The regular summation theorem~\cite{Luscher:1986pf}, which is used in order
to perform the continuum limit in the sums over the discrete momentum
 eigenvalues, implies that the integrand is a continuous function in the 
momenta. However, the finite-volume 
matrix elements $\langle 0|\phi(0)|\beta\rangle$, which
enter Eq.~(\ref{eq:analytic}), contain L\"uscher's zeta-function and can become
singular. Here, for  one particular example, we shall demonstrate how these
singularities are lifted.

The averaged quantities, for which the validity of the regular
summation theorem will be checked, are defined in Eq.~(\ref{eq:fromappendix}).
From now on, without loss of generality, we shall work in the center-of-mass
frame ${\bf k}=0$. We wish to demonstrate that
\eq
\lim_{L\to\infty}A(\omega_0,\Delta,{\bf 0})= A^\infty(\omega_0,\Delta,{\bf
  0})\, ,\quad\quad \omega_0>\omega_{\sf min}\, ,\quad\Delta>0\, .
\en
More precisely, the difference between the both sides of the above equation
vanishes faster that any negative power of $L$, as $L\to\infty$.

Let $\omega_0$ be in the elastic scattering region. Since $L$ is large, 
characteristic momenta are small and non-relativistic quantum mechanics
provides an adequate description of a problem under consideration.
 Let us consider two massive (distinguishable) particles in the CM frame.
The state vector corresponding to the eigenvalue $E_\beta$ is given by
\eq
|\beta\rangle=\frac{1}{L^{d/2}}\,\sum_{\bf q} f_\beta({\bf q})\,
|{\bf q},-{\bf q}\rangle\, ,\quad\quad
|{\bf q},-{\bf q}\rangle=a_1^\dagger({\bf q})a_2^\dagger(-{\bf q})|0\rangle\, ,
\en
where $a_i^\dagger,\, i=1,2$ denote the creation operators for the particles
1 and 2, respectively, and the wave function $f_\beta({\bf q})$ is normalized,
according to
\eq
\label{eq:norm}
\langle\beta|\beta\rangle=\frac{1}{L^d}\,\sum_{\bf q}
|f_\beta({\bf q})|^2=1\, .
\en
For simplicity, let us further assume that the interaction between the particles 
is described by a separable potential 
$V({\bf p},{\bf k})=gv({\bf p})v({\bf k})$, where the function
$v({\bf p})$ corresponds to a 
smooth cutoff at large momenta. Note that in the following we will never need 
the explicit form of this function. The $T$-matrix is given by
\eq
T({\bf p},{\bf q};E)=\frac{v({\bf p})v({\bf q})}{g^{-1}-I(E)}\, ,\quad\quad
I(E)=\frac{1}{L^d}\,\sum_{\bf k}\frac{v^2({\bf k})}{{\bf k}^2-k_0^2}
\, , \qquad 
k_0^2=2\mu E\, ,
\en
where $\mu$ denotes the reduced mass of the system.

In the limit $E\to E_\beta$ the $T$-matrix has a pole. In the vicinity of the
pole, it behaves as
\eq
T({\bf p},{\bf q};E)=\frac{v({\bf p})v({\bf q})}{g^{-1}-I(E_\beta)
-I'(E_\beta)(E-E_\beta)+\cdots}
=\frac{v({\bf p})v({\bf q})}{-I'(E_\beta)(E-E_\beta)+\cdots}\, .
\en
From this expression, we may read off the wave function corresponding
to the eigenvalue $E_\beta$
\eq
f_\beta({\bf q})=\sqrt{\frac{2\mu}{I'(E_\beta)}}\,\frac{v({\bf q})}
{{\bf q}^2-k_{0\beta}^2}\, ,\quad\quad
k_{0\beta}^2=2\mu E_\beta\, .
\en
The normalization of this wave function was chosen so that $f_\beta({\bf p})$
obeys Eq.~(\ref{eq:norm}).

Take the composite field $\phi(0)=\phi_1(0)\phi_2(0)$, where
the $\phi_i(0),\,i=1,2$ denote the elementary particle fields. In 
momentum space,
\eq
\phi_i(0,{\bf x})
=\frac{1}{L^{d/2}}\,\sum_{\bf k} \mbox{e}^{i{\bf k}{\bf x}}\,a_i({\bf k})
\en
The matrix element
that enters the spectral function is given by
\eq\label{eq:matrix}
\langle 0|\phi(0)|\beta\rangle
=\frac{1}{L^{3d/2}}\,\sum_{\bf q}f_\beta({\bf q})
\doteq \frac{1}{L^d}\,\tilde f_\beta({\bf 0})\, ,
\en
where $\tilde f_\beta({\bf r})$ denotes the Fourier-transform 
of $f_\beta({\bf q})$.
The averaged spectral function is written in the following form
\eq
A(\omega_0,\Delta,{\bf 0})=\frac{1}{L^d}\,
\sum_\beta\theta(\omega_0+\Delta/2-E_\beta)\theta(E_\beta-\omega_0+\Delta/2)\,
|\tilde f_\beta({\bf 0})|^2\, .
\en
Hence, in order to verify the applicability of the regular summation theorem in
this case, it suffices to show that $\tilde f_\beta({\bf 0})$ is a regular
function of $k_{0\beta}$. As anticipated, this function contains L\"uscher's
zeta-function which is singular at $k_{0\beta}^2={(2\pi{\bf n})^2}/{L^2}$.
However, the factor $I'(E_\beta)$, which enters the normalization, contains
the zeta-function as well. It is easy to check that the singular factors
in the numerator and the denominator cancel, and the regular summation 
theorem holds.

\section{Emergence of the second Riemann sheet in the
infinite volume limit}
\label{app:Herglotz}

Let us consider\footnote{We are indebted to J\"urg Gasser who indicated
  this example to us.} the function $F_L$ of the complex variable $z$
\eq
F_L(z)=\frac{1}{1-z+\epsilon^2\sqrt{z}\cot(\sqrt{z}L)}\, ,\quad
\quad z\in\mathbb{C}\, .
\en
Note that this function resembles the propagator of an unstable particle in
the 1+1-dimensional effective field theory~\cite{Hoja2}. Further,
it can be shown that
\eq
{\rm sign}({\rm Im}\,(\sqrt{z}\cot(\sqrt{z}L)))=-{\rm sign}({\rm Im}\,(z))\, .
\en
According to this condition, the denominator can not vanish outside the real axis. Thus,
the only singularities of $F_L(z)$ are simple poles on the positive real
axis.   

If ${\rm Im}\,(z)\neq 0$, in the limit $L\to\infty$ we have
$\cot(\sqrt{z}L)\to -i\,{\rm sign}({\rm Im}\,(z))$ and, therefore,
\eq
F_\infty(z)=\frac{1}{1-z+\epsilon^2\sqrt{-z}}\, .
\en
The difference $|F_L(z)-F_\infty(z)|$ vanishes exponentially with $L$, if
${\rm Im}\,(z)\neq 0$. Note that, unlike $F_L(z)$, which is a meromorphic
function, $F_\infty(z)$ is analytic in the complex plane cut along the
positive real axis. This is what is meant when we speak of   
``the poles merging into the cut''.

Moreover, the function $F_\infty(z)$ has a couple of complex-conjugated poles
{\em on the second Riemann sheet}. These poles are solutions of the equation
$1-z+\epsilon^2\sqrt{-z}$ that gives $z_{\pm}=1\mp i\epsilon^2+O(\epsilon^4)$.
If $\epsilon^2$ is small, these poles come close to the physical
scattering region and influence  $F_\infty(z)$ on the physical sheet. Since
away from the real axis the difference between   $F_\infty(z)$ and $F_L(z)$
vanishes exponentially at a large $L$, the effect of the 
poles on the second Riemann sheet
is felt in  $F_L(z)$ as well.

\end{document}